\documentclass[iop,tighten,twocolumn,preprintnumbers]{emulateapj}
\usepackage{lineno}
\modulolinenumbers[5]
\usepackage{amsmath,bbold,amssymb,epsfig,bm,mathrsfs,feynmp}
\usepackage{color,slashed,nicefrac,exscale,multirow,soul}
\usepackage{times,txfonts,xcolor,graphicx,gensymb,tikz}
\usepackage[normalem]{ulem}
\usepackage[breaklinks,colorlinks,citecolor=blue]{hyperref}
\usepackage{eso-pic}
\newcommand{\reportnum}[2]{
  \AddToShipoutPictureBG*{%
    \AtPageUpperLeft{%
      \hspace{0.75\paperwidth}%
      \raisebox{#1\baselineskip}{%
        \makebox[0pt][l]{\textnormal{#2}}
  }}}%
}
\usepackage{lineno}
\definecolor{red}{rgb}{0.8,0,0}
\definecolor{violet}{rgb}{0.4,0,0.4}
\definecolor{green}{rgb}{0,0.5,0.0}
\definecolor{navy}{rgb}{0.0,0.0,0.6}
\definecolor{orange}{rgb}{0.8,0.2,0.0}

\newcommand{\blue}[1]{\textcolor[rgb]{0.00,0.00,1.00}{#1}}

\newcommand{\jjl}[1]{{\color[rgb]{0.4,0.0,0.4}#1}}

\usepackage[normalem]{ulem}  

\newcommand{\bea}{\begin{eqnarray}}
\newcommand{\eea}{\end{eqnarray}}
\newcommand{\ep}{\varepsilon}

\begin{document}
\reportnum{-8}{INT-PUB-23-029}
\title{New covariant density functionals of nuclear matter\\ 
for compact star simulations\vspace{-1.5cm}}
\author{Jia Jie Li$^1$ and Armen Sedrakian$^{2,3}$}
\affiliation{
    $^1$School of Physical Science and Technology, Southwest University, Chongqing 400715, China;
    \blue{jiajieli@swu.edu.cn}\\
$^2$Frankfurt Institute for Advanced Studies,
    D-60438 Frankfurt am Main, Germany;\\
$^3$Institute of Theoretical Physics,
    University of Wroclaw, 50-204 Wroclaw, Poland\\
}
\begin{abstract}
We generate three families of extended covariant density functionals
of nuclear matter that have varying slope of symmetry energy and
skewness at nuclear saturation density, but otherwise share the same 
basic parameters (symmetry  energy, compressibility, saturation parameters, 
etc.) with the standard DDME2, DD2, and MPE functionals. Tables of the 
parameters of these new density functionals are given, which can be 
straightforwardly used in DDME2, DD2, and  MPE parameterization-based 
codes. Furthermore, we  provide tables of a large number of equations of 
state (81 for each family) that can be used in astrophysical simulations 
to assess the impact of variations of not-well-known slope of symmetry 
energy and skewness of nuclear systems on the astrophysics of compact objects. We also provide tables of computed integral parameters (mass, radius, and 
tidal deformability) that can be used, e.g., for modeling gravitational 
waveforms. Finally, for the extended DDME2-based parameterization, we 
implement a first-order phase transition to quark matter to obtain a 
family of equations of state that accommodates a phase transition 
to quark matter. Analogous tables of the equations of state and integral parameters 
are provided for this case as well. 
\end{abstract}
\maketitle
%
\section{Introduction}
\label{sec:Intro}
The detection of gravitational waves from binary compact sources 
by first- and second-generation gravitational-wave detectors and the
prospect of higher-sensitivity  observations by the third-generation 
gravitational observatories, such as the Einstein Telescope, provide 
a strong motivation for modeling the astrophysical sources of gravitational 
waves. Binary neutron star  mergers offer significant information 
via the inference of the tidal deformation in the premerger phase, as 
already observed in the GW170817 
event~\citep{LIGO-Virgo:2017,LIGO-Virgo:2018,LIGO-Virgo:2019}, 
as well as through observations of oscillations of the postmerger 
objects~\citep{LVC:2017,LVC:2019}. 

An important input in binary neutron star modeling is the equation of state
(EoS) of dense matter in compact stars. To date, a large number of 
EoSs have been proposed which can be classified in the following groups: 
(a)~the microscopic EoS which are based on first-principle computations, 
such as those based on potentials fitted to the nucleon–nucleon scattering 
data or those based on QCD, either in the perturbative or low-energy 
limits~\citep[for recent reviews, see][]{Oertel:2017,Burgio:2021,Drischler:2019}; 
(b)~density functional-based EoSs, which start with some form of energy
density functional with adjustable parameters that are then fitted to
bulk properties of nuclei and/or nuclear 
matter~\citep[for recent reviews, see][]{Oertel:2017,BlaschkeChamel:2018,Sedrakian:2023}; 
(c) matter-composition-agnostic models, which are based on some form 
of a parameterization of the EoS, well-known examples being 
polytropes~\citep{Ozel:2009,Read:2009}, constant speed of sound (CSS)
parameterizations~\citep{Alford:2013,Zdunik:2013}, metamodels based 
on Taylor expansions of energy density close to saturation 
density~\citep{Margueron:2018,Ferreira:2022,BoanLi:2023}, and non-parametric
EoSs~\citep{Landry:2019,Landry:2020,Essick:2021}, etc.

The covariant density functionals (CDFs) provide a fast and reliable
method to incorporate physical constraints from bulk properties of
nuclear systems and astrophysical information starting from 
baryon-meson Lagrangians~\citep{Serot:1997,Bender:2003,Vretenar:2005,Mengj:2006,Oertel:2017}. 
They provide access to microscopic information, such as self-energies 
(decomposed according to their Lorentz structure), matter composition, 
chemical potentials, and effective masses. At the same time, they allow 
for flexible adjustments of model parameters with the inflow of new 
information and constraints on the EoS or other aspects of micro- and 
macro-physics. A good example is how CDF-based theories confront the 
so-called ``hyperon puzzle"; see~\cite{Sedrakian:2023} and references 
therein.

To obtain a CDF, the partition function of a many-body system is often
evaluated in the lowest-order Hartree approximation, which leads to
relativistic mean-field 
models~\citep{Serot:1997,Bender:2003,Vretenar:2005,Mengj:2006,Oertel:2017}. 
However, those CDFs that reproduce the available data to high accuracy 
do not strictly represent mean-field realizations of any given Lagrangian, 
as the correlations beyond the mean-field are built into the fitting procedure 
via additional elements, such as 
(a) nonlinear mesonic fields~\citep[e.g.,][]{Boguta:1977,Lalazissis:1997,Todd-Rutel:2005,Chenwc:2014} 
and (b) the density dependence of the coupling constants~\citep[e.g.,][]{Typel:1999,Long:2004,Lalazissis:2005,Typel:2010,Typel:2018}. 
As a result, the self-energies are not linear functions of density, as 
is the case in the mean-field theory. For this reason, we will avoid 
the term ``relativistic mean-field model" and use CDF theory instead.

In this work, we provide a large set of EoSs that are based on CDF
descriptions of nuclear matter. Our point of departure will be three standard CDFs with density-dependent couplings given
by~\cite{Lalazissis:2005}, \cite{Typel:2010} and \cite{Typel:2018}, which
were fitted to the laboratory data (commonly, the binding energies,
charge radii, spin-orbit splittings, the neutron skin thickness,
etc. for a set of magic and semimagic nuclei) using a least-squares
minimization procedure. Note that the fitting procedures are not
identical, but are based on the calibration of parameters on a particular set of the physical observables of finite nuclei and/or nuclear matter.
Next, we generate a large number of CDFs by varying the
density dependence of the coupling constants. These variations are
then mapped onto the  variations of the coefficients of the expansion 
of the energy density of isospin asymmetric nuclear matter in Taylor 
series around the nuclear saturation and isospin symmetrical limits. 
As a proof of principle, we have previously provided a smaller sample 
of CDF EoSs~\citep{Lijj:2019b}, which were mapped on the EoS models
following from purely phenomenological expansions of the nuclear
matter EoS near the saturation density that is most prominently 
used in the metamodels of EoSs~\citep{Margueron:2018}, but also in the
Bayesian analysis of EoS 
parameters~\citep{Char:2020,Traversi:2020,Malik:2023,Zhuzy:2023} and 
for providing sensible constraints on parameters of nuclear and neutron
star matter~\citep{Lattimer:2004,Lattimer:2023}.
Thus, the first goal of this work is the construction of three new and 
large families of nucleonic EoSs that at saturation reproduce a given
standard CDF, such as DDME2~\citep{Lalazissis:2005}, DD2~\citep{Typel:2010}, 
and MPE~\citep{Typel:2018}. They, however, possess an alternative 
density dependence of the couplings and, thus, the resulting coefficients of 
Taylor expansion of energy density. These families are constructed to 
cover a substantial range of the mass-radius diagram of compact stars 
that is compatible with current astrophysical constraints. This will 
allow us to carry out systematic studies of the dependence of 
astrophysical simulations on the important not-well-determined parameters 
of the nuclear EoS, such as the slope of symmetry energy and skewness of 
symmetric matter at nuclear saturation density. The second goal extends 
the DDME2 functional-based analysis to allow for a phase transition to 
quark matter. The quark matter EoS is modeled using simple CSS 
parameterization~\citep{Alford:2013,Zdunik:2013}.
We assume a strong, first-order phase transition to quark matter in 
this case. Thus, this particular set of EoSs allows astrophysical
implementations of QCD phase transition in dense matter. Finally,
all the generated data for the EoSs and  the integral parameters of compact
stars, such as mass, radius, and tidal deformability, are provided in
the form of tables, that can be downloaded from the 
https://github.com/asedrakian/DD\_CDFs  
repository, while static 
versions are available in Zenodo at doi:10.5281/zenodo.8350680.
This work is structured as follows. In Section~\ref{sec:Construction} we
briefly review the formalism of CDF theory at the saturation density
and its standard realizations in the case of density-dependent couplings. 
In Section~\ref{sec:NM} we propose new CDFs with modified couplings and 
present their predictions for the EoS, the mass-radius diagram, and 
the tidal deformabilities of static, spherically symmetrical stars. 
Section~\ref{sec:Hybrid} describes the case where a first-order 
phase transition to quark matter is allowed for. Here we present the 
same parameters of static hybrid stars as in the case of nucleonic models. 
Our conclusions are collected in Section~\ref{sec:Conclusions}.

\section{Covariant Density Functionals}
\label{sec:Construction}
In this work, we use the CDF approach based on the Lagrangian of
stellar matter with nucleonic degrees of freedom
$ \mathscr{L} = \mathscr{L}_N + \mathscr{L}_m + \mathscr{L}_l +
\mathscr{L}_{\rm em}, $
where the nucleonic Lagrangian is given by
\begin{eqnarray}
\label{eq:Lagrangian}
\mathscr{L}_N  = 
\sum_N\bar\psi_N \Big[\gamma^\mu
\big(i \partial_\mu - g_{\omega}\omega_\mu 
    - g_{\rho} \bm{\tau} \cdot \bm{\rho}_\mu \big)
    -\big(m_N - g_{\sigma}\sigma\big) \Big]\psi_N, 
\nonumber \\
\end{eqnarray}
where $\psi_N$ are the nucleonic Dirac fields with masses $m_N$, 
and $\sigma,\,\omega_\mu$, and $\bm{\rho}_\mu$ are the mesonic fields 
that mediate the interaction among nucleon fields. The remaining pieces 
of the Lagrangian correspond to the mesonic, leptonic, and 
electromagnetic contributions, respectively~\citep{Sedrakian:2023}. 

The meson-nucleon couplings which are density-dependent and 
are given by~\citep{Typel:1999,Typel:2018}
\begin{align}
g_{m}(\rho)=g_{m}(\rho_{\rm {ref}})f_m(r),
\end{align}
with a constant value $g_{m}(\rho_{\rm{ref}})$ at a reference 
density $\rho_{\rm{ref}}$, and a function $f_m(r)$ that depends 
on the ratio $r=\rho/\rho_{\rm {ref}}$. For the isoscalar channel, 
one has a rational function
\begin{align}\label{eq:isoscalar_coupling}
f_{m}(r)=a_m\frac{1+b_m(r+d_m)^2}{1+c_m(r+d_m)^2}, \quad m = \sigma,\omega,
\end{align}
with conditions
\begin{align}\label{eq:coupling_constraints}
f_{m}(1)=1, \quad f^{\prime\prime}_{m}(0)=0, \quad
f^{\prime\prime}_{\sigma}(1)=f^{\prime\prime}_{\omega}(1),
\end{align}
which reduce the number of free parameters.
The density dependence for the isovector channel is 
taken in an exponential form:
\begin{align}\label{eq:isovector_coupling}
f_{m}(r) = e^{-a_m (r-1)}, \quad m = \rho.
\end{align}

\begin{table}[b]
\caption{Nuclear matter characteristics at saturation density 
of the reference CDF parameterizations.} 
\setlength{\tabcolsep}{4.0pt}
\label{tab:Nuclear_matter}
\begin{tabular}{cccccccccccc}
\hline\hline
CDF    &$\rho_{\rm sat}$ &$E_{\rm sat}$&$K_{\rm sat}$&$Q_{\rm sat}$&$J_{\rm sym}$&$L_{\rm sym}$&$M^\ast_D$\\
       &[fm$^{-3}$]&[MeV]    &[MeV]    &[MeV]        &[MeV]        &[MeV]        & [$m_N$]\\
\hline
DDME2  & 0.1520 & $-16.14$ & 251.15 &479.22 & 32.31 & 51.27 & 0.57\\
DD2    & 0.1491 & $-16.02$ & 242.72 &169.15 & 31.37 & 55.04 & 0.56\\
MPE    & 0.1510 & $-16.10$ & 242.79 &223.14 & 32.51 & 58.42 & 0.58\\
\hline\hline
\end{tabular}
\tablecomments{$M^\ast_D$ denotes for the average Dirac effective mass 
in units of nucleon mass.}
\end{table}

The energy density of isospin asymmetric matter is customarily split into an isoscalar and an isocector term:
\begin{eqnarray}
\label{eq:isospin_expansion}
E(\rho, \delta) \simeq 
E_0(\rho) + E_{\rm sym}(\rho)\delta^2 + {\mathcal O}(\delta^4)
\end{eqnarray}
where $\rho= \rho_n + \rho_p$ is the baryonic density, with $\rho_{n(p)}$ denoting the neutron (proton) density, $\delta = (\rho_{n}-\rho_{p})/\rho$ is the isospin asymmetry, and $E_0(\rho)$ and $E_{\rm sym}(\rho)$ are, respectively, the energy of symmetric matter and the symmetry energy. At densities close to 
the saturation $\rho_{\rm sat}$ Equation~\eqref{eq:isospin_expansion} can be further Taylor expanded as
\begin{eqnarray}
\label{eq:Taylor_expansion}
E(\rho, \delta) 
&\simeq&
E_{\rm{sat}} + \frac{1}{2!}K_{\rm{sat}}\chi^2
+ \frac{1}{3!}Q_{\rm{sat}}\chi^3 \nonumber \\ [1.0ex]
&  & +\,J_{\rm{sym}}\delta^2 + L_{\rm{sym}}\delta^2\chi + K_{\rm{sym}}\delta^2\chi^2
+ {\mathcal O}(\chi^4, \chi^3\delta^2),
\nonumber \\
\end{eqnarray}
where $\chi=(\rho-\rho_{\rm{sat}})/3\rho_{\rm{sat}}$. The coefficients of the expansion are known as
{\it nuclear matter characteristics at saturation density}, namely,
{\it incompressibility} $K_{\rm{sat}}$, the {\it skewness} $Q_{\rm{sat}}$, the {\it symmetry energy} $J_{\rm{sym}}$ and its 
{\it slope parameter} $L_{\rm{sym}}$. The coefficients of the
leading-order terms in the expansion, namely $K_{\rm{sat}}$ and
$J_{\rm{sym}}$ are well determined, whereas the coefficients of higher-order terms, namely $Q_{\rm{sat}}$ and $L_{\rm{sym}}$ are not well
known. By definition, these coefficients allow one to assess the properties of nuclear matter predicted from various models.

In the most widely-used CDF parameterizations, e.g.,
DDME2~\citep{Lalazissis:2005} and DD2~\citep{Typel:2010}, the
reference density is chosen as the vector density at saturation
$\rho_{\rm {ref}} = \rho^{v}_{\rm sat}$. If we fix in the
Lagrangian~\eqref{eq:Lagrangian} the nucleon and meson masses, the
properties of infinite nuclear matter (the \jjl{seven} characteristics listed in
Table~\ref{tab:Nuclear_matter}) can be computed uniquely in terms of
seven adjustable parameters~\citep{Lijj:2019b}. These are three
coupling constants at saturation density
($g_{\sigma},\,g_{\omega},\,g_{\rho}$) and four parameters
($a_\sigma,\,b_\sigma,\,a_\omega,\,a_\rho$) that control their density
dependence. Besides these choices, we also use the newly proposed MPE
parameterization~\citep{Typel:2018}, for which dependence on the
scalar density is considered for the coupling of scalar meson $\sigma$
and the reference density $\rho_{\rm {ref}} = \rho^{s}_{\rm sat}$. 
In this case, the third condition in Eq.~\eqref{eq:coupling_constraints} is dropped.  The predicted nuclear matter parameters with these three CDFs 
are summarized in Table~\ref{tab:Nuclear_matter}. It is evident that the choice of the reference density is part of the definition of 
the CDF and thus characterizes it.

\section{Nucleonic models}
\label{sec:NM}
In this work, we use the DDME2, DD2, and MPE CDF parameterizations 
as reference CDFs on which to build our extension. 
Because the nuclear 
matter coefficients $J_{\rm sym}$ and $L_{\rm sym}$ are strongly 
correlated while the value of symmetry energy $E_{\rm sym}$ at 
density $\rho_c = 0.11$~fm$^{-3}$ is almost identical for a variety of 
nuclear models~\citep[e.g.,][]{Trippa:2008,Ducoin:2011,Lijj:2018a}, 
we hold the value of $E_{\rm{sym}}(\rho_c)$ constant for each family 
of CDF when $L_{\rm sym}$ is being varied. Consequently, $J_{\rm sym}$ varies as well. There are also correlations between other pairs of coefficients. For example, $K_{\rm sat}$-$Q_{\rm sat}$ pair is 
correlated but to a less noticeable 
degree~\citep[e.g.,][]{Typel:2018,Malik:2020}. A correlation among 
the pair $L_{\rm sym}$-$K_{\rm sym}$ may also 
exist~\citep[e.g.,][]{Ducoin:2011,Providencia:2013}, such correlations being a property of CDF itself. 
Having fixed the values of 
$\rho_{\rm{sat}}$, $E_{\rm{sat}}$, $K_{\rm{sat}}$, and 
$E_{\rm{sym}}(\rho_c)$, we proceed to quantify the uncertainties in 
the CDFs by covering a range of terms of not-well-known higher-order 
characteristics of nuclear matter, specifically $Q_{\rm sat}$ and 
$L_{\rm sym}$. 
To this end, for isoscalar $\sigma$- and $\omega$-mesons we fix their values of couplings at saturation density, but modify their density dependences; while for isovector $\rho$-meson we fix its coupling value at $\rho_c$ but modify its density dependence, so the value of coupling at saturation density is thus changed accordingly. The parameters entering each set of nuclear matter coefficients are obtained by a numerical search.

We then compute the EoS of cold stellar matter by implementing the 
conditions of $\beta$-equilibrium and charge neutrality,
where both 
electrons and muons are included. We further match smoothly our core 
EoS to that of inner crust EoS given in~\cite{Baym:1971a} at 
the crust-core transition density $\sim \rho_{\rm sat}/2$. For the outer crust, we use the EoS given in~\cite{Baym:1971b}.
With this input, the integral parameters of compact stars, in particular, 
the mass $M$ and the radius $R$, are then computed from the 
Tolman-Oppenheimer-Volkoff (TOV) equations~\citep{Tolman:1939,Oppenheimer:1939}.
The tidal deformability $\lambda$ critically characterizes 
how easily an object can be deformed under an external tidal field~\citep{Hinderer:2007,Binnington:2009}. It is given as 
$\lambda = 2/3\,k_2 R^{5}$, where $k_2$ is the dimensionless Love number 
that can be computed simultaneously with the TOV equations. We work with the dimensionless $\Lambda = \lambda M^{-5}$ for convenience.

\begin{figure}[tb]
\centering
\includegraphics[width = 0.47\textwidth]{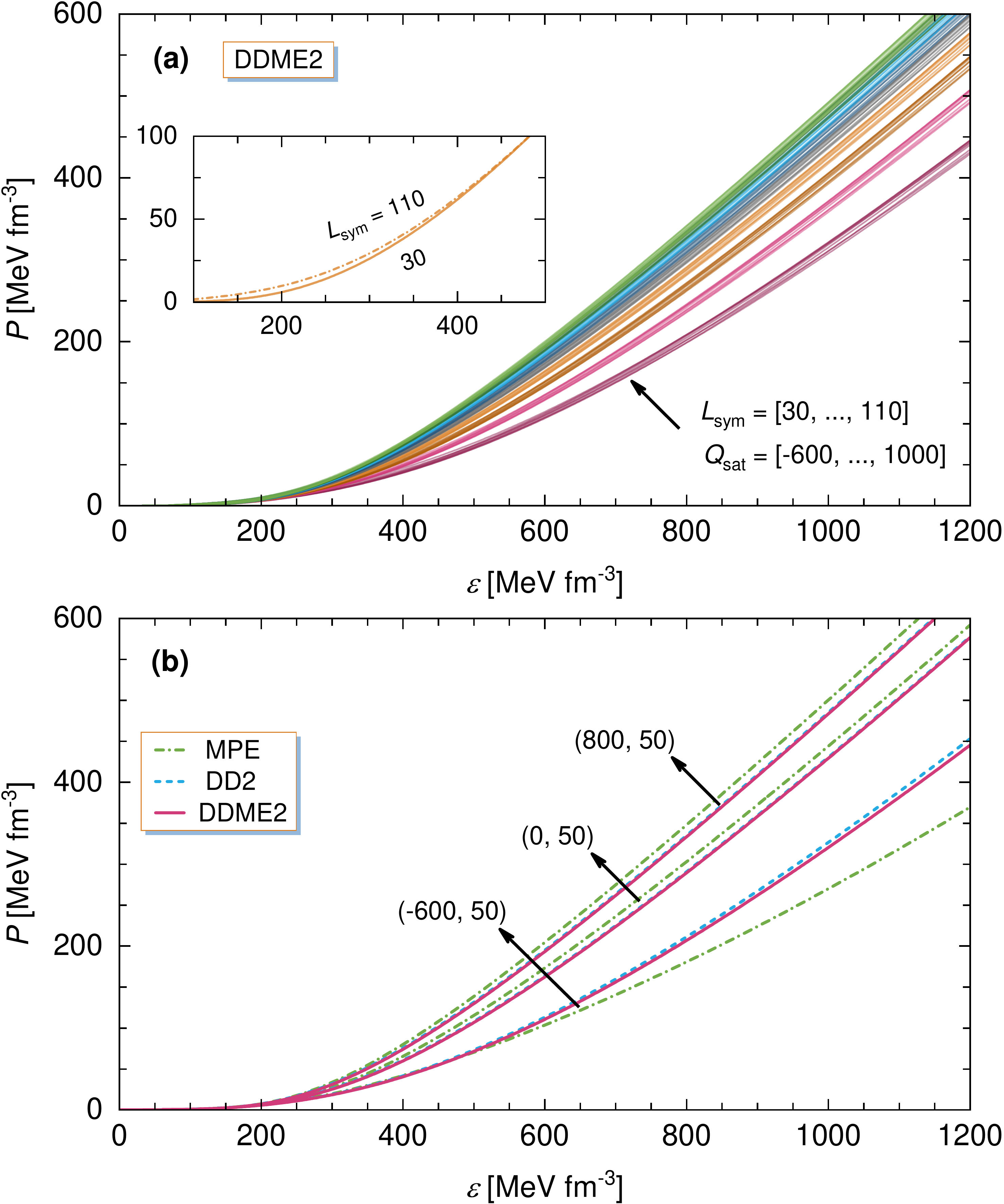}
\caption{EoS for purely nucleonic stellar matter. 
In panel (a) the models are generated with DDME2 family of CDF models 
by varying the parameters $Q_{\rm sat} \in [-600,\,1000]$~MeV and 
$L_{\rm sym} \in [30,\,110]$~MeV. The effects of parameter $L_{\rm sym}$ 
on the low-density region of EoS are shown in the inset for illustration. 
In panel (b) the same is shown for three families of CDF models with 
fixed values of pairs $(Q_{\rm sat},\,L_{\rm sym})$ (in MeV) as 
indicated in the plot.}
\label{fig:EoS_QL}
\end{figure}

Figure~\ref{fig:EoS_QL} shows the EoS of stellar matter generated 
by DDME2 family of CDF models [panel (a)] and for three families of CDF 
models, DDME2, DD2 and MPE, for pairs of $(Q_{\rm sat},\,L_{\rm sym})$ 
values as indicated, thus showing the impact of the form of density 
functional on the EoS [panel (b)]. 

Our collection in Figure~\ref{fig:EoS_QL}~(a) contains in total 81 EoSs for 
each fixed set of low-order characteristics of DDME2, which are generated 
by varying $-600 \le Q_{\rm sat} \le 1000$~MeV and 
$30 \le L_{\rm sym} \le 110$~MeV
with steps $\Delta Q_{\rm sat} =200$~MeV and $\Delta L_{\rm sym} =10$~MeV.
Tables~\ref{tab:DDME-1} and~\ref{tab:DDME-2} in Appendix provide the
corresponding parameter sets for each CDF. The range of these parameters 
is chosen as follows. For $Q_{\rm sat} = -600$~MeV and  independent of
the value $L_{\rm sym}$ the maximum mass is about $2.0\,M_\odot$, which 
matches the mass measurement of PSR 
J0740+6620~\citep{NANOGrav:2019,Fonseca:2021}. 
For $Q_{\rm sat} \geq 600$~MeV, the maximum mass of a compact star reaches 
the value $2.59^{+0.08}_{-0.09}\,M_\odot$ deduced for the light companion 
of GW190814~\citep{LIGO-Virgo:2020} if interpreted as a neutron star. 
The lower and upper values of $L_{\rm sym}$ are those predicted by the
majority of the density functionals and are within the range currently 
discussed~\citep[see, e.g.,][]{PREX-II:2021,CREX:2022,Lattimer:2023,Sedrakian:2023}.

\begin{figure}[tb]
\centering
\includegraphics[width = 0.47\textwidth]{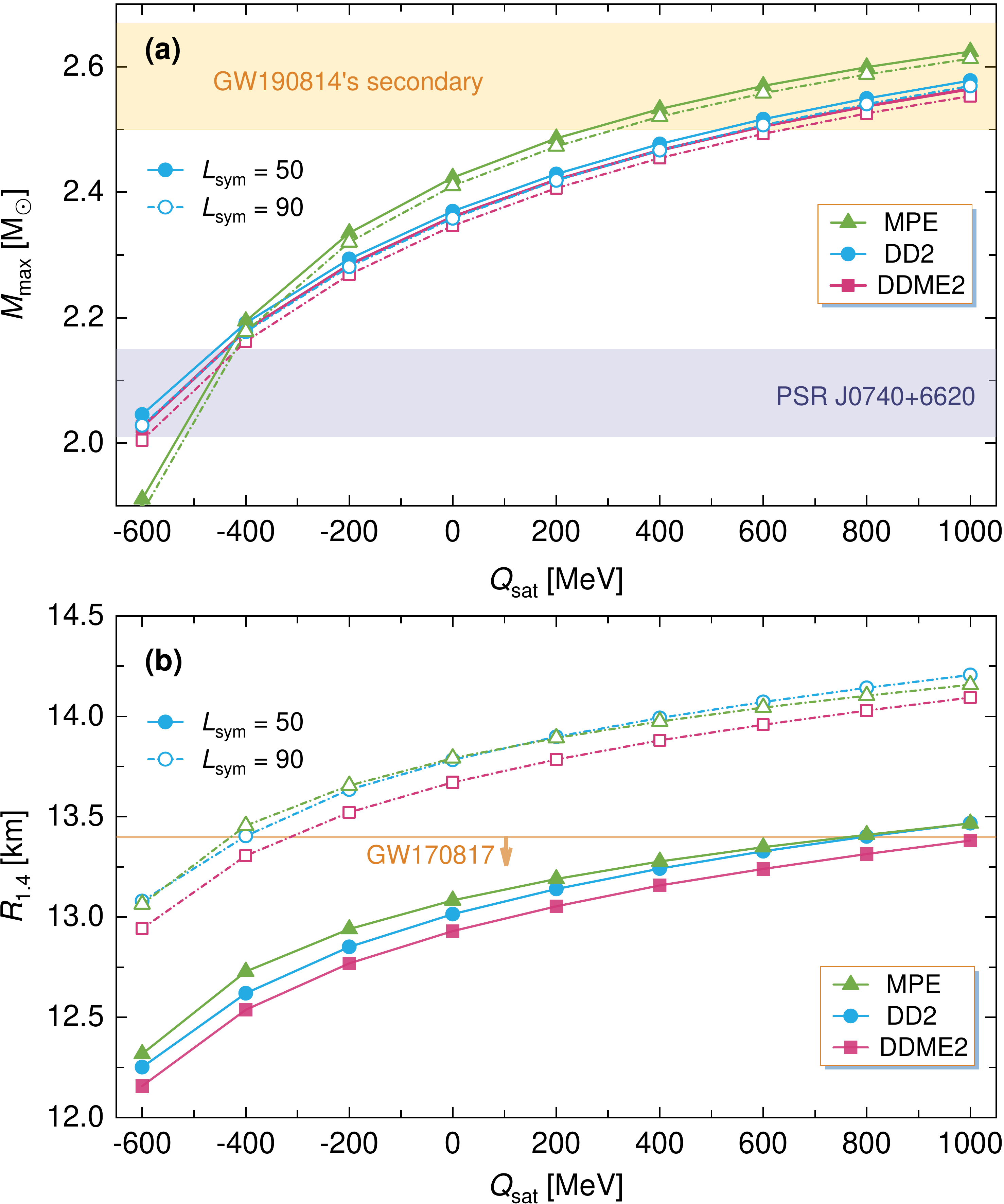}
\caption{
The maximum mass $M_{\rm max}$ [panel (a)] and the radius $R_{1.4}$ 
of a canonical 1.4 $M_{\odot}$  mass star [panel (b)] that are predicted 
by the three families of EoSs generated from each CDF used for various values 
of $L_{\rm sym}$ as a function of $Q_{\rm sat}$ (in MeV). In panel (a) 
the shadings show the masses of 
PSR~J0470+6620~\citep{NANOGrav:2019,Fonseca:2021} and the secondary 
component of GW190814~\citep{LIGO-Virgo:2020}, while in panel (b), 
the vertical line indicates the upper limit on the radius set by the 
analysis of GW170817~\citep{LIGO-Virgo:2018}. 
}
\label{fig:MR_QL}
\end{figure}

Next, let us assess the influences of  the low-order characteristics and
the different functional forms of the couplings that we 
have fixed within 
each family of CDFs on the EOS and the gross properties of neutron stars.
From Figure~\ref{fig:EoS_QL}~(b), where we vary the form of the CDF by 
fixing pairs of values $(Q_{\rm sat},\,L_{\rm sym})$, we observe that:
(a) the differences in the low-order characteristics of nuclear matter 
among the three families of CDFs, which control the properties of finite 
nuclei and nuclear matter at subsaturation densities, have negligible 
impact on the EoS, especially at high densities, which are more relevant 
for neutron stars. This is clearly seen from the comparison of the DDME2 and 
DD2 CDFs which become identical for the stiff EoS in each family;
and 
(b) the high-density behavior of the EoS depends strongly on the  
specific chosen family of CDFs (which differ by their functional form) and/or 
the choice of density dependencies of couplings on scalar or vector density. 
As seen from the comparison of the DDME2 and MPE CDFs, sizable differences 
emerge in the high-density domain, especially for those parameters that 
lead to soft EoSs.

Figure~\ref{fig:MR_QL} shows the maximum mass $M_{\rm max}$ [panel (a)] 
and the radius $R_{1.4}$ of a canonical $1.4\,M_\odot$ mass star [panel (b)] 
for static, spherically symmetrical stellar configurations predicted by the 
three families of EoSs generated from the three different CDFs we used. Here 
we vary continuously $Q_{\rm sat}$ for some representative values of $L_{\rm sym}$. 
In agreement with the observed difference among EoSs, it is clearly seen that 
the choice of the specific form of the CDF and corresponding low-order 
characteristics has a small impact ($< 2\%$) on the radius of canonical 
$1.4\,M_{\odot}$ mass star. The variations of such a magnitude expected for 
the low-mass domain of compact stars are significantly smaller than current measurement accuracy of $\sim 10\%$. At the same time, the differences in the maximum masses predicted by DDME2/DD2 and MPE CDFs
can reach up to 5\%, which correspond to $\sim 0.1\,M_\odot$, and are comparable to the error bars in the mass measurements of very massive (close or beyond two solar mass) compact stars. Therefore, the choice of the functional form of the CDF (among many other factors) influences the predictions for the properties of compact stars and justifies considering different families of CDFs in astrophysical studies of compact stars.

\begin{figure}[tb]
\centering
\includegraphics[width = 0.47\textwidth]{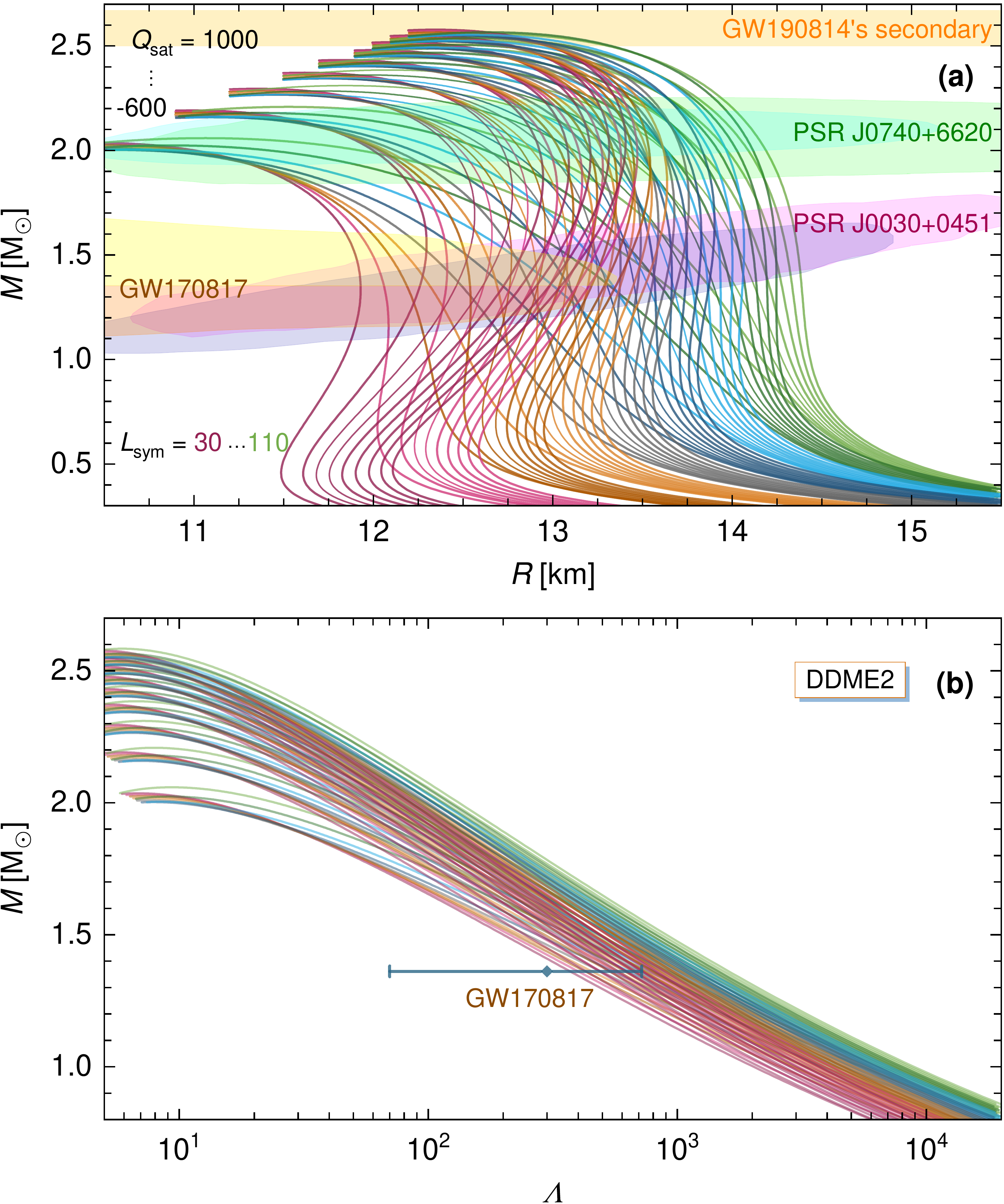}
\caption{Mass-radius [panel (a)] and mass-tidal deformability [panel (b)] 
relations for nucleonic EoS models with different pairs of values of 
$Q_{\rm sat}$ and $L_{\rm sym}$ (in MeV). In panel (a) the color regions 
show the 90\% confidence interval (CI) ellipses from each of the two NICER modeling groups for 
PSR~J0030+0451 and J0740+6620~\citep{NICER:2019a,NICER:2021a,NICER:2019b,NICER:2021b},
the 90\% CI regions for each of the two compact stars that merged in the 
gravitational-wave event GW170817~\citep{LIGO-Virgo:2019}, and finally
the 90\% CI for the mass of the secondary component of 
GW190814~\citep{LIGO-Virgo:2020}. In panel (b), the constraint for a 
$1.36\,M_{\odot}$ star deduced from the analysis of GW170817
event~\citep{LIGO-Virgo:2019} is shown too.
}
\label{fig:MR_ML}
\end{figure}

Figure~\ref{fig:MR_ML} shows the mass-radius curves [panel (a)] and 
mass-tidal deformability relations [panel (b)] for static stellar 
configurations based on the 81 EoS of the DDME2 family. Constraints 
from multimessenger astronomy are highlighted for comparison. These 
concern the mass and radius measurements for 
PSR~J0030+0451~\citep{NICER:2019a,NICER:2019b} and 
J0740+6620~\citep{NICER:2021a,NICER:2021b} by NICER, the compactness 
and tidal deformability constraints extracted from the binary compact star 
merger GW170817~\citep{LIGO-Virgo:2019}, and the mass of the secondary 
component of GW190814 event~\citep{LIGO-Virgo:2020}. For these observations, 
all the regions/limits are given at 90\% credible intervals. As seen from 
Figure~\ref{fig:MR_ML}, the softness of the EoSs at the intermediate density 
implied by the GW170817 event and the stiffness at high density suggested by 
the massive pulsars/objects can be reconciled by an appropriate choice of 
the parameters of $L_{\rm sym}$ and $Q_{\rm sat}$.

Similarly, we generate the collection from the 
MPE family, for which the
parameters are provided in Tables~\ref{tab:MPE-1} and~\ref{tab:MPE-2} in the 
Appendix. Note that with this functional form, a slightly higher value 
$Q_{\rm sat} \geq -500$\,MeV is required to reach the high mass 
$2.08^{+0.07}_{-0.07}\,M_{\odot}$ of PSR~J0740+6620~\citep{NANOGrav:2019,Fonseca:2021}; 
see Figure~\ref{fig:MR_QL}~(a).

\section{Hybrid models}
\label{sec:Hybrid}
\begin{figure}[!]
\centering
\includegraphics[width = 0.47\textwidth]{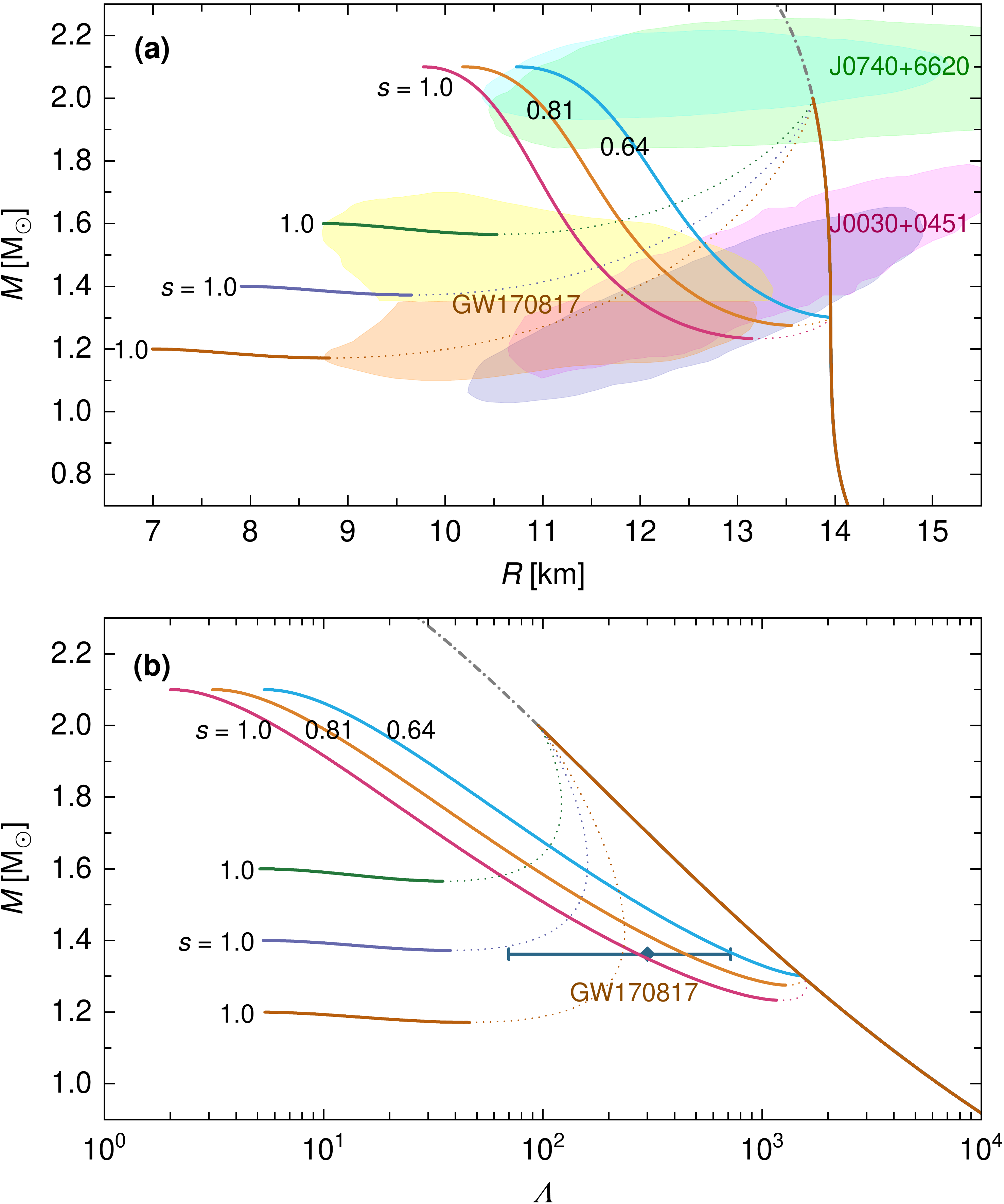}
\caption{Mass-radius [panel (a)] and mass-tidal deformability 
[panel (b)] relations for hybrid EoS models constructed from a 
stiff nucleonic EoS ($Q_{\rm sat} = 600$,\,$L_{\rm sym} = 90$~MeV)
with phase transitions at low and high densities. For all the hybrid 
branches, the values of sound speed squared $c^2_s$ are as indicated  
in the plot. The dotted thin lines indicate unstable configurations. 
Constraints from multimessenger astronomy are shown too.
}
\label{fig:MR_Hybrid}
\end{figure}

About half of our models from the previous section have tidal
deformabilities that are outside the inference for GW170817. 
However, these EoSs can be reconciled with this inference if a 
phase transition to quark matter occurs. In this section, we 
demonstrate the  utility of such models by supplementing them 
with a quark EoS using as a model the CSS
parameterization~\citep{Zdunik:2013,Alford:2013}
\bea\label{eq:EoS}
p(\ep) =\left\{
\begin{array}{ll}
p_{\rm tran}, \quad & \ep_{\rm tran} < \ep < \ep_{\rm tran}\!+\!\Delta\ep, \\[0.5ex]
p_{\rm tran} + c^2_s\,\bigl[\ep-(\ep_{\rm tran}\!+\!\Delta\ep)\bigr],
  \quad & \ep_{\rm tran}\!+\!\Delta\ep < \ep, 
\end{array} 
\right.\nonumber \\
\eea
where $p_{\rm tran}$ is the transitional pressure with energy density
$\ep_{\rm tran}$ at nucleonic density $\rho_{\rm tran}$, $\Delta\ep$ is 
the energy discontinuity, and $c_s$ is the sound speed in the quark phase. 

Figure~\ref{fig:MR_Hybrid} illustrates how such a transition to quark
matter allows a nuclear EoS with a large $L_{\rm sym}$ to be consistent 
with astrophysical constraints. We show two scenarios in which the phase 
transition occurs at low or high enough densities, respectively, 
considering different values of $c_s$. For stiff quark EoS ($c_s \approx 1.0$), 
which allows a large enough jump in energy density $\Delta\ep$ under 
consideration, there can be a second stable branch of compact stars. 
This leads to the emergence of twin configurations --- two stable 
compact stars have an identical mass, but different radii and, consequently, 
tidal deformabilities~\citep[e.g.,][]{Alford:2013,Benic:2015,Paschalidis:2018,Alvarez-Castillo:2019,Christian:2019,Montana:2019}. In the first scenario, the 
phase transition occurs at $\rho_{\rm tran}/\rho_{\rm sat} \lesssim 2.0$,
causing the nucleonic branch to end at a lower mass $\lesssim 1.4\,M_{\odot}$, 
and the hybrid branch extends up to 
$M_{\rm max} > 2\,M_{\odot}$~\citep[e.g.,][]{Paschalidis:2018,Montana:2019,Blaschke:2020,Lijj:2021,Christian:2022,Lijj:2023a}.
In the second scenario, in contrast, the phase transition occurs at 
$\rho_{\rm tran}/\rho_{\rm sat} \gtrsim 3.0$, so that the heaviest 
star on the nucleonic branch has $M_{\rm max} > 2\,M_{\odot}$, 
and the hybrid branches lie at lower masses, which yield 
ultracompact  hybrid stars~\citep{Lijj:2023b}. 

Note that ellipses/limits referring to GW170817 have been obtained under 
the assumption of nucleonic stars, therefore they are relevant, strictly 
speaking, only for constraining the nucleonic EoS models. The ultracompact 
hybrid star models shown in Figure~\ref{fig:MR_Hybrid}, however, could satisfy 
the mutual dependence of tidal deformabilities $\Lambda_1-\Lambda_2$ of 
members of a binary for GW170817~\citep{Lijj:2023b}. In this scenario, 
the multimessenger data can be accounted for if PSR~J0030+0451 and 
J0740+6620 are purely nuclear stars with large radii, whereas one of the 
components of GW170817 binary is ultracompact hybrid star~\citep{Lijj:2023b}. 
Ultracompact star arise in scenarios where the conformal limit of QCD 
is reached at quite high densities. If this density is low, then the 
mass-radius curves are limited to the regions with larger radii, around 
10-12 km~\citep{Sedrakianparticles:2023}.

\section{Conclusions}
\label{sec:Conclusions}
In this work, we constructed three families of CDFs starting from 
DDME2, DD2, and MPE parameterization allowing for variations of the slope 
of the symmetry energy and skewness of symmetric matter at saturation 
density. The new CDFs allow for systematic studies of the sensitivity 
of various astrophysical scenarios toward variations of these parameters. 
One way to utilize the results is to modify the codes that are based on 
the original CDFs with density-dependent couplings to generate the EoS 
and other observables. Tables~\ref{tab:DDME-2} and \ref{tab:MPE-2} in the
Appendix provide the parameter values of extended density functionals 
for such purpose. Another way to utilize our results is to use directly 
the provided tables of EoSs in numerical simulations. In addition, we 
provided tables of integral parameters, such as the mass, radius, and 
tidal deformability, which can be used in astrophysical analysis, for 
example, to generate gravitational-wave templates. The effects of 
first-order phase transition in neutron stars were studied using the 
simple CSS parameterization of the EoS of quark matter within two 
scenarios of low- and high-density phase transition to quark matter. 
The EoS and integral parameter tables are also provided in this case.

So far, we have concerned only been concerned with purely nucleonic models. Of course, 
 hyperonization and the onset of resonances are a possibility in compact 
stars, in which case both the EoS and integral parameters are 
modified~\citep[e.g.,][]{Chatterjee:2016,Tolos:2020,Sedrakian:2023}. 
Clearly, our alternative parameterizations can be used to study 
hypernuclear matter, eventually admixed with $\Delta$-resonances, 
to study the sensitivity of the EoS on the high-order characteristics 
of nuclear matter.

\section*{Acknowledgements}
J.~L. acknowledges the support of the National Natural Science 
Foundation of China (grant No. 12105232), the Fundamental Research 
Funds for the Central Universities (Grant No. SWU-020021), and 
the Venture \& Innovation Support Program for Chongqing Overseas 
Returnees (grant No. CX2021007). A.~S. acknowledges the support 
from the Deutsche  Forschungsgemeinschaft grant No. SE 1836/5-2, 
the Polish NCN  grant No. 2020/37/B/ST9/01937 at Wroc\l{}aw University, 
and the Institute for Nuclear Theory at the University of Washington, for 
its kind hospitality, stimulating research environment, and partial 
support from the INT's U.S. Department of Energy grant No. DE-FG02-00ER41132.

\appendix
\section{Parameter values of the DDME2 and MPE families}
In this appendix we present the values of the parameters determining 
the CDFs. The parameters of the original DDME2 and MPE parameterizations 
are shown in Tables~\ref{tab:DDME-1} and \ref{tab:MPE-1}, respectively. 
As we already discussed in the text, these models are well constrained 
with respect to the lower-order saturation characteristics. In
Tables~\ref{tab:DDME-2} and~\ref{tab:MPE-2} we further present sets 
of alternative parameterizations that produce different values of
$Q_{\rm{sat}}$ and/or $L_{\rm{sym}}$, but preserve these 
lower-order 
values predicted by the original DDME2 and MPE parameterizations.
To this end, one needs to modify only the parameters in functions 
$f_{\sigma}$ and $f_{\omega}$ [see Equation~\eqref{eq:isoscalar_coupling}] 
which control the density dependence of the couplings in the isoscalar 
sector, and $f_{\rho}$ [see Equation~\eqref{eq:isovector_coupling}] in the 
isovector sector.

\begin{table*}[htb]
\centering
\caption{Nucleon, meson masses and meson-nucleon coupling 
  constants in the DDME2 parameteriztation, where $g_{m}$ 
  ($m = \sigma$, $\omega$, and $\rho$) refer to the values 
  at the saturation density.}
\setlength{\tabcolsep}{13.8pt}
\label{tab:DDME-1}
\centering
\begin{tabular}{cccccccccc}
\hline\hline
$m_n$      & $m_p$      & $m_\sigma$ & $m_\omega$ & $m_\rho$   & $g_{\sigma}$ &$g_{\omega}$ & $g_{\rho}$ &         \\
939.0000   & 939.0000   &  550.1238  &  783.0000  & 763.0000   &   10.5396      &    13.0189    &   3.6836  &         \\
\hline
$a_\sigma$ & $b_\sigma$ & $c_\sigma$ & $d_\sigma$ & $a_\omega$ & $b_\omega$     & $c_\omega$    & $d_\omega$   & $a_\rho$\\
  1.3881   &   1.0943   &   1.7057   &   0.4421   &   1.3892   &   0.9240       &    1.4620  
 &    0.4775    &  0.5647 \\
\hline\hline
\end{tabular}
\tablecomments{The masses are in units of MeV.}
\end{table*}

\begin{table*}[htb]
\centering
\caption{Alternative parameterization of the density dependence of the
  couplings in the isoscalar and isovector channels for the indicated values of $Q_{\rm{sat}}$ (MeV) and/or $L_{\rm{sym}}$ (MeV). The values of $g_{\sigma}$ and $g_{\omega}$ are the same as in the 
  DDME2 parametrization that given at the saturation density, while 
  that  for $g_{\rho}$ here refers to the value at the cross density 
  $\rho_c = 0.11$~fm$^{-3}$.
  }
\setlength{\tabcolsep}{7.6pt}
\label{tab:DDME-2}
\centering
\begin{tabular}{cccccccccccc}
\hline\hline
$Q_{\rm{sat}}$ & $L_{\rm{sym}}$ & $a_{\sigma}$ & $b_{\sigma}$ & $c_{\sigma}$ & $d_{\sigma}$ & $a_{\omega}$ & 
$b_{\omega}$ & $c_{\omega}$ & $d_{\omega}$ & $g_{\rho}$ & $a_{\rho}$ \\
\hline
 -600 & 30-110 & 1.3489 & 0.1774 & 0.3256 & 1.0119 & 1.3775 & 0.1449 & 0.2870 & 1.0777 & 4.2970 & 0.8550, 0.6926, 0.5562 \\
      &        &        &        &        &        &        &        &        &        &        & 0.4380, 0.3333, 0.2391 \\
      &        &        &        &        &        &        &        &        &        &        & 0.1532, 0.0742, 0.0011 \\[0.6ex]     
 -400 & 30-110 & 1.3401 & 0.2773 & 0.4720 & 0.8404 & 1.3630 & 0.2299 & 0.4143 & 0.8970 & 4.2993 & 0.8629, 0.6992, 0.5620 \\
      &        &        &        &        &        &        &        &        &        &        & 0.4433, 0.3381, 0.2435 \\
      &        &        &        &        &        &        &        &        &        &        & 0.1574, 0.0781, 0.0048 \\[0.6ex]
 -200 & 30-110 & 1.3397 & 0.4004 & 0.6519 & 0.7151 & 1.3577 & 0.3350 & 0.5696 & 0.7650 & 4.3012 & 0.8698, 0.7051, 0.5671 \\
      &        &        &        &        &        &        &        &        &        &        & 0.4478, 0.3423, 0.2474 \\
      &        &        &        &        &        &        &        &        &        &        & 0.1610, 0.0815, 0.0080 \\[0.6ex]
  0   & 30-100 & 1.3459 & 0.5526 & 0.8760 & 0.6169 & 1.3593 & 0.4646 & 0.7616 & 0.6616 & 4.3027 & 0.8761, 0.7103, 0.5717 \\
      &        &        &        &        &        &        &        &        &        &        & 0.4519, 0.3460, 0.2508 \\
      &        &        &        &        &        &        &        &        &        &        & 0.1641, 0.0844, 0.0108 \\[0.6ex]    
  200 & 30-100 & 1.3583 & 0.7420 & 1.1598 & 0.5361 & 1.3670 & 0.6255 & 1.0028 & 0.5766 & 4.3040 & 0.8818, 0.7150, 0.5758 \\
      &        &        &        &        &        &        &        &        &        &        & 0.4556, 0.3493, 0.2538 \\
      &        &        &        &        &        &        &        &        &        &        & 0.1670, 0.0872, 0.0134 \\[0.6ex]    
  400 & 30-100 & 1.3777 & 0.9806 & 1.5264 & 0.4673 & 1.3812 & 0.8274 & 1.3114 & 0.5042 & 4.3051 & 0.8871, 0.7194, 0.5796 \\
      &        &        &        &        &        &        &        &        &        &        & 0.4589, 0.3524, 0.2567 \\
      &        &        &        &        &        &        &        &        &        &        & 0.1696, 0.0897, 0.0157 \\[0.6ex]    
  600 & 30-110 & 1.4057 & 1.2856 & 2.0121 & 0.4070 & 1.4030 & 1.0843 & 1.7155 & 0.4408 & 4.3060 & 0.8920, 0.7235, 0.5831 \\
      &        &        &        &        &        &        &        &        &        &        & 0.4621, 0.3552, 0.2592 \\
      &        &        &        &        &        &        &        &        &        &        & 0.1720, 0.0919, 0.0178 \\[0.6ex]     
  800 & 30-110 & 1.4455 & 1.6829 & 2.6760 & 0.3529 & 1.4348 & 1.4171 & 2.2603 & 0.3840 & 4.3067 & 0.8966, 0.7273, 0.5864 \\
      &        &        &        &        &        &        &        &        &        &        & 0.4650, 0.3578, 0.2616 \\
      &        &        &        &        &        &        &        &        &        &        & 0.1743, 0.0940, 0.0198 \\[0.6ex]
 1000 & 30-110 & 1.5025 & 2.2122 & 3.6196 & 0.3035 & 1.4806 & 1.8572 & 3.0206 & 0.3322 & 4.3073 & 0.9010, 0.7309, 0.5895 \\
      &        &        &        &        &        &        &        &        &        &        & 0.4677, 0.3603, 0.2639 \\
      &        &        &        &        &        &        &        &        &        &        & 0.1763, 0.0959, 0.0216 \\
\hline\hline
\end{tabular}
\tablecomments{For $\rho$ meson coupling, the value of $g_\rho (\rho_{\rm sat})$ 
can be simply obtained through formula 
$g_\rho (\rho_{\rm sat}) = g_\rho (\rho_c) e^{-a_{\rho} (1 - \rho_c/\rho_{\rm sat})}$, 
and a larger value of $a_{\rho}$ corresponds to
a lower value of the slope parameter $L_{\rm sym}$.}  
\end{table*}

\begin{table*}[htb]
\centering
\caption{Nucleon, meson masses and meson-nucleon coupling constants 
  in the MPE parametriztation, whereby $g_{m}$ ($m = \sigma$, $\omega$, 
  and $\rho$)refer to the values at the saturation density, and 
  $\rho^s_{\rm sat}$ the corresponding scalar density.}
\setlength{\tabcolsep}{13.8pt}
\label{tab:MPE-1}
\centering
\begin{tabular}{cccccccccc}
\hline\hline
$m_n$      & $m_p$      & $m_\sigma$ & $m_\omega$ & $m_\rho$   & $g_{\sigma}$ &$g_{\omega}$ & $g_{\rho}$ &  $\rho^s_{\rm sat}$   \\
939.5654   & 938.2721   &  559.1141  &  783.0000  & 763.0000   &   11.1052    &    13.5784    &   3.7625     &   0.1419  \\
\hline
$a_\sigma$ & $b_\sigma$ & $c_\sigma$ & $d_\sigma$ & $a_\omega$ & $b_\omega$   & $c_\omega$    & $d_\omega$   & $a_\rho$\\
  1.2078   &   0.4421   &   0.6023   &   0.7439   &   1.1936   &   0.1958     &    0.2778   
& 1.0954   &  0.4877 \\
\hline\hline
\end{tabular}
\tablecomments{The masses are in units of MeV, and density in fm$^{-3}$.}
\end{table*}

\begin{table*}[htb]
\centering
\caption{Alternative parametrization of the density dependence of the
  couplings in the isoscalar and isovector channels for the indicated 
  values of $Q_{\rm{sat}}$ (MeV) and/or $L_{\rm{sym}}$ (MeV). 
  The values of $g_{\sigma}$ and $g_{\omega}$ are the same as in the 
  MPE parametrization that given at the saturation density, while  
  that for $g_{\rho}$ here refers to the value at the cross density 
  $\rho_c = 0.11$~fm$^{-3}$.}
\setlength{\tabcolsep}{7.6pt}
\label{tab:MPE-2}
\centering
\begin{tabular}{cccccccccccc}
\hline\hline
$Q_{\rm{sat}}$ & $L_{\rm{sym}}$ & $a_{\sigma}$ & $b_{\sigma}$ & $c_{\sigma}$ & $d_{\sigma}$ & $a_{\omega}$ & 
$b_{\omega}$ & $c_{\omega}$ & $d_{\omega}$ & $g_{\rho}$ & $a_{\rho}$ \\
\hline
 -600 & 30-110 & 1.1976 & 0.1368 & 0.2016 & 1.2858 & 1.2521 & 0.0212 & 0.0446 & 2.7341 & 4.2894 & 0.8859, 0.7180, 0.5776 \\
      &        &        &        &        &        &        &        &        &        &        & 0.4562, 0.3488, 0.2523 \\
      &        &        &        &        &        &        &        &        &        &        & 0.1644, 0.0837, 0.0089 \\[0.6ex]
 -400 & 30-110 & 1.1958 & 0.1929 & 0.2750 & 1.1009 & 1.2169 & 0.0509 & 0.0867 & 1.9604 & 4.2907 & 0.8907, 0.7221, 0.5811 \\
      &        &        &        &        &        &        &        &        &        &        & 0.4594, 0.3517, 0.2549 \\
      &        &        &        &        &        &        &        &        &        &        & 0.1669, 0.0859, 0.0111 \\[0.6ex]
 -200 & 30-110 & 1.1973 & 0.2595 & 0.3621 & 0.9595 & 1.2022 & 0.0890 & 0.1380 & 1.5542 & 4.2921 & 0.8955, 0.7260, 0.5846 \\
      &        &        &        &        &        &        &        &        &        &        & 0.4625, 0.3545, 0.2576 \\
      &        &        &        &        &        &        &        &        &        &        & 0.1694, 0.0883, 0.0133 \\[0.6ex]
  0   & 30-100 & 1.2011 & 0.3381 & 0.4651 & 0.8466 & 1.1957 & 0.1348 & 0.1983 & 1.2966 & 4.2935 & 0.9004, 0.7302, 0.5882 \\
      &        &        &        &        &        &        &        &        &        &        & 0.4657, 0.3575, 0.2603 \\
      &        &        &        &        &        &        &        &        &        &        & 0.1719, 0.0907, 0.0156 \\[0.6ex]
  200 & 30-100 & 1.2073 & 0.4302 & 0.5868 & 0.7537 & 1.1940 & 0.1896 & 0.2699 & 1.1113 & 4.2950 & 0.9052, 0.7342, 0.5917 \\
      &        &        &        &        &        &        &        &        &        &        & 0.4689, 0.3604, 0.2630 \\
      &        &        &        &        &        &        &        &        &        &        & 0.1744, 0.0931, 0.0178 \\[0.6ex]
  400 & 30-100 & 1.2151 & 0.5386 & 0.7311 & 0.6752 & 1.1946 & 0.2525 & 0.3516 & 0.9736 & 4.2966 & 0.9105, 0.7386, 0.5956 \\
      &        &        &        &        &        &        &        &        &        &        & 0.4723, 0.3635, 0.2659 \\
      &        &        &        &        &        &        &        &        &        &        & 0.1771, 0.0956, 0.0203 \\[0.6ex]
  600 & 30-110 & 1.2254 & 0.6659 & 0.9031 & 0.6075 & 1.1978 & 0.3264 & 0.4480 & 0.8626 & 4.2983 & 0.9156, 0.7428, 0.5992 \\
      &        &        &        &        &        &        &        &        &        &        & 0.4756, 0.3665, 0.2687 \\
      &        &        &        &        &        &        &        &        &        &        & 0.1798, 0.0981, 0.0227 \\[0.6ex]
  800 & 30-110 & 1.2382 & 0.8154 & 1.1090 & 0.5482 & 1.2032 & 0.4125 & 0.5611 & 0.7707 & 4.3001 & 0.9205, 0.7470, 0.6029 \\
      &        &        &        &        &        &        &        &        &        &        & 0.4789, 0.3695, 0.2715 \\
      &        &        &        &        &        &        &        &        &        &        & 0.1824, 0.1006, 0.0250 \\[0.6ex]
 1000 & 30-110 & 1.2538 & 0.9923 & 1.3576 & 0.4955 & 1.2104 & 0.5121 & 0.6931 & 0.6935 & 4.3020 & 0.9255, 0.7512, 0.6066 \\
      &        &        &        &        &        &        &        &        &        &        & 0.4822, 0.3726, 0.2743 \\
      &        &        &        &        &        &        &        &        &        &        & 0.1851, 0.1031, 0.0274 \\
\hline\hline
\end{tabular}
\tablecomments{For $\rho$ meson coupling, the value of $g_\rho (\rho_{\rm sat})$ 
can be simply obtained through formula 
$g_\rho (\rho_{\rm sat}) = g_\rho (\rho_c) e^{-a_{\rho} (1 - \rho_c/\rho_{\rm sat})}$, 
and a larger value of $a_{\rho}$ corresponds to
a lower value of the slope parameter $L_{\rm sym}$.}  
\end{table*}

\newpage

\end{document}